# Design and Analysis of a Spurious Switching Suppression Technique Equipped Low Power Multiplier with Hybrid Encoding Scheme

S.Saravanan
*Department of ECE,*
*K.S.R.College of Technology*
Tiruchengode-637215, India. .

M.Madheswaran
*Department of ECE*
*Muthayammal Engineering College*
Rasipuram-647408, India

*Abstract—* **Multiplication is an arithmetic operation that is mostly used in Digital Signal Processing (DSP) and communication applications. Efficient implementation of the multipliers is required in many applications. The design and analysis of Spurious Switching Suppression Technique (SSST) equipped low power multiplier with hybrid encoding is presented in this paper. The proposed encoding technique reduces the number of switching activity and dynamic power consumption by analyzing the bit patterns in the input data. In this proposed encoding scheme, the operation is executed depends upon the number of 1's and its position in the multiplier data. The architecture of the proposed multiplier is designed using a low power full adder which consumes less power than the other adder architectures. The switching activity of the proposed multiplier has been reduced by 86% and 46% compared with conventional and Booth multiplier respectively. It is observed from the device level simulation using TANNER 12.6 EDA that the power consumption of the proposed multiplier has been reduced by 87% and 26% compared with conventional and Booth multiplier**

## I. INTRODUCTION

Multiplication is one of the most critical operations in many computational systems. The growing popularity of portable and multimedia devices such as video phones, note books in which multipliers play the important role. This has motivated the research in the recent years to design low power VLSI circuits. Application specific integrated circuits rely on efficient implementation of various arithmetic circuits for executing the specified algorithms. It is well known that if the density of transistor increases, the complexity of arithmetic circuits also increases and consumes more energy. This has further motivated the new concepts of designing low power VLSI circuits. It is also clear that the reduction in power consumption and enhancement in the circuit design are expected to pose challenges in implementing wireless multimedia and digital image processing system, in which multiplication and multiplication-accumulation are the key computations. In the recent past, the researchers proposed various design methodologies on dynamic power reduction using minimizing the switching activities. Choi et al [1] proposed Partially Guarded Computation (PGC) which divides the arithmetic units into two parts and turns off the unused part to minimize the power consumption. The reported results show that the PGC can reduce power consumption by 10% to 44% in an array multiplier with 30% to 36% area overhead in speech related applications. A 32-bit 2's complement adder equipping a dynamic-range determination (DRD) unit and a sign-extension unit was reported by Chen et al [2]. This design tends to reduce the power dissipation of conventional adders for multimedia applications. Later, Chen et al [3] presented a multiplier using the DRD unit to select the input operand with a smaller effective dynamic range that yield the Booth codes which reduces 30% power dissipation than conventional method. Benini et al [4] reported the technique for glitching power minimization by replacing few existing gates with functionally equivalent ones that can be "frozen" by asserting a control signal. This method operates in the layout level environment which is tightly restricted and hence it reduces 6.3% of total power dissipation. The double-switch circuit-block scheme was proposed by Henzler et al [5] is capable of reducing power dissipation by shortening the settling time during down time. Huang and Ercegovac [6] presented the arithmetic details about the signal gating schemes and showed 10% to 45% power reduction for adders. The combination of the signal flow optimization (SFO), left-to-right leapfrog (LRLF) structure, and upper/lower split structure was incorporated in the design to optimize the array multipliers by Huang and Ercegovac [7] and it is reported that the new approach can save about 20% power dissipation.

Wen et al [8] reported that for the known output, some columns in the multiplier can be turned off and reduce 10% power consumption for random inputs. Chen and Chu [9] later, reported that the spurious power suppression technique has been applied on both compression tree and modified Booth decoder to enlarge the power reduction. Ko et al [10] and Song and Micheli [11] investigated full adder as the core element of complex arithmetic units like adder, multiplier, division, exponentiation and MAC units. Several combinations of static CMOS logic styles have been used to implement low-power one bit adder cells. In general, the logic styles were broadly divided into two major categories such as the complementary CMOS and the pass-transistor logic circuits. The complementary CMOS logic style uses the power lines as input where the pass transistor logic uses separate input signals.

The complementary CMOS full adder is based on the regular CMOS structure with pMOS pull-up and nMOS pull-down transistors [12]. The authors reported that the series transistors in the output stage form a weak driver and additional buffers at the last stage is required for providing the necessary driving power to the cascaded cells. Chandrakasan and Brodersen [13] reported that the Complementary Pass transistor Logic (CPL) full adder with swing restoration structure utilizes 32 transistors. A Transmission Function Full Adder (TFA) based on the transmission function theory was presented by Zhuang and Hu [14]. Later, Weste and Eshraghian [15] presented a Transmission Gate Adder (TGA) using CMOS transmission gates circuit which is a special kind of pass-transistor logic circuit. The transmission gate logic requires double the number of transistors of the standard pass-transistor logic or more to implement the same circuit. Hence the research has been focused by various researchers on smaller transistor count adder circuits, most of which exploit the non full swing pass transistors with swing restored transmission gate technique. This is exemplified by the state-of-the-art design of 14T and 10T which was reported by Vesterbacka [16] and Bui et al [17]. Chang et al [18] proposed a hybrid style full adder circuit in which the sum and carry generation circuits are designed using hybrid logic styles.

Full adders are used in a tree structure for high performance arithmetic circuits and a cascaded simulation structure is introduced to evaluate the full adders in real time applications. Keeping the above facts, it is proposed to improve the performance of the multiplier unit using spurious switching suppression technique and hybrid encoded scheme. In this research paper a novel design method has been proposed to reduce the number of switching activities and power consumption of multiplier.

## II. PROPOSED HYBRID ENCODED LOW POWER MULTIPLIER

### A. Hybrid encoding rule

In general, multiplication process consists of two parts as multiplicand and multiplier. According to the conventional shift and add multiplication, the number of partial products (PP) are equal to the number of bits in the multiplier. The number of partial products can be reduced by half using Booth recoding. In the proposed encoding technique, the partial products can still be reduced which in turn reduces the switching activity and power consumption. The operation can be defined according to the number of 1's and its position in the multiplier. The proposed hybrid encoding rule is demonstrated and provided in Fig 1. The operation of proposed hybrid encoding rule is stated in Table 1 with details of operation.

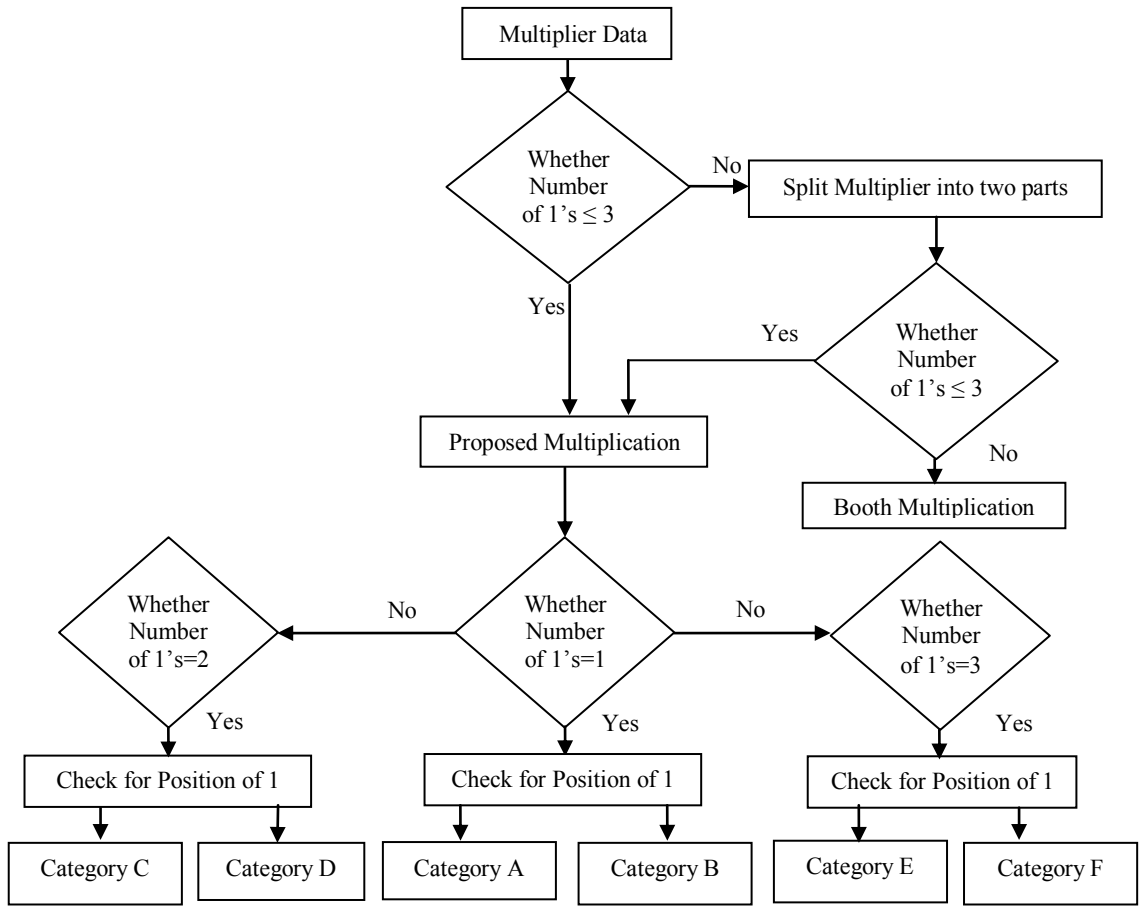

Figure 1.   Flow chart of the proposed multiplier

If the number of 1's in the multiplier is less than or equal to 3, the control goes to proposed multiplication technique, otherwise the control split the multiplier in to two parts. Again the number of 1's in the part of the multiplier is verified. If the number of 1's is more than three, the control goes to Booth multiplication. Otherwise the control goes to proposed multiplication technique. If the number of 1's in the multiplier is one and depends upon its position, the control goes to execute the operation in category A or B. If the number of 1's in the multiplier is two and depends upon its position, the control goes to execute the operation in category C or D. Otherwise the number of 1's in the multiplier is three and depends upon its position, the control goes to execute the operation in category E or F. The proposed multiplication technique is explained with the example shown in Fig. 2.

TABLE I.    HYBRID ENCODING SCHEME

| Number of 1's in the Multiplier | Position of the 1 | Category | Operation |
|---|---|---|---|
| 1 | $1^{st}$ bit | A | Add 0 to multiplicand (M) |
| 1 | $i^{th}$ bit | B | Shift M left by i-1 and add 0 |
| 2 | $1^{st}$ and $i^{th}$ bit | C | Shift M left by i-1 and add M |
| 2 | $i^{th}$ and $i+j^{th}$ bit | D | Shift M left by j , add M and shift the result left by i-1 |
| 3 | $i^{th}=1^{st}$, $j^{th}$ and $k^{th}$ bit | E | Shift M by k-j , add M and shift the result left by j-i, add M and shift the result left  by i-1 |
| 3 | $i^{th}$, $j^{th}$ and $k^{th}$ bit | F | Shift M by k-j , add M and shift the result left by j-i, add M and shift the result left  by i |

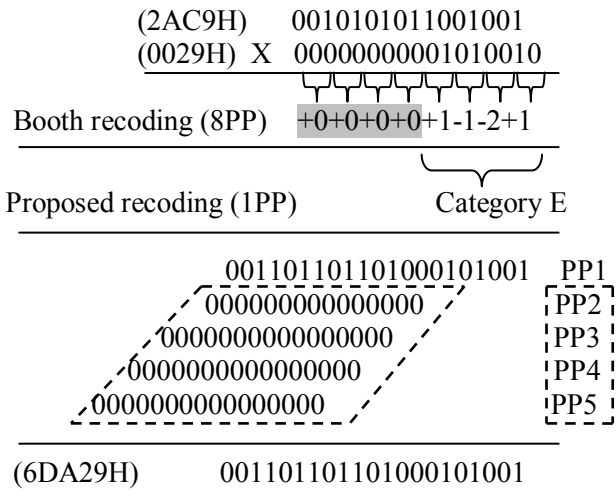

Figure 2.    Proposed hybrid encoded multiplication

For the above multiplication, conventional multiplication scheme needs 16 partial products and 15 addition operations, Booth multiplication needs 8 partial products and 7 addition operation.

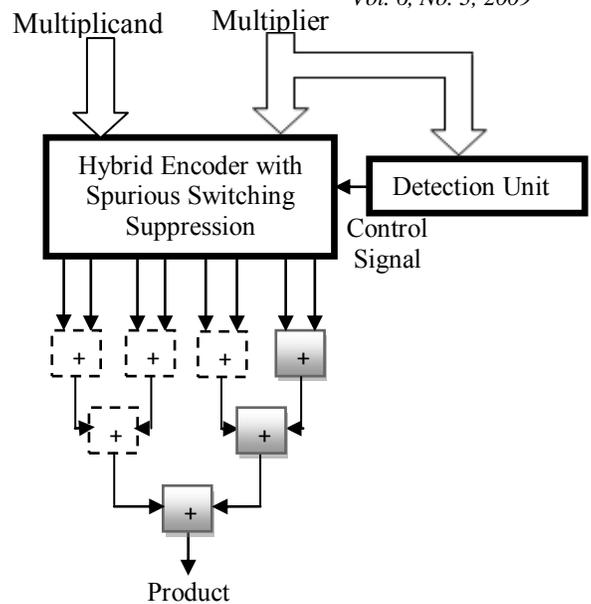

Figure 3.    SSST equipped multiplier

According to category E, the proposed encoding rule needs one partial product P1 with two additions. The remaining partial products P2 to P5 are zero, so the addition operation in this area can be neglected, which reduces the switching activity and power consumption. This spurious switching activity can be reduced by freezing the adders which perform this unwanted addition. In Fig 3 dashed adders operation can be blocked and shadowed adders only functioning to avoid the unwanted switching operation and power.

### B.  Block diagram of proposed hybrid encoded low power multiplier

The block diagram of the proposed hybrid encoded low power multiplier is shown in Fig 4. The process of the proposed multiplier can be divided into hybrid encoding, multiplication and controlling. The proposed encoder works as per the method explained in Fig 1. In the partial product compression the partial products are added without carry propagation and row bypassing can be used when the entire row of the PP is zero. This is expected to reduce the switching activity and power consumption. In the final adder unit a column bypassing provision is available to avoid the unwanted addition operation. The detection logic circuit is used to detect the effective data range. If the part of the input data does not make any impact in the final computing results then the data controlling circuit freezes that portion to avoid unnecessary switching transitions. A glue circuit can be used to control the carry and sign extension unit which will manage the sign.

### III.   RESULTS AND DISCUSSIONS

The low power multiplier circuit which is a part of MAC unit has been simulated using TANNER 12.6 EDA schematic editor.

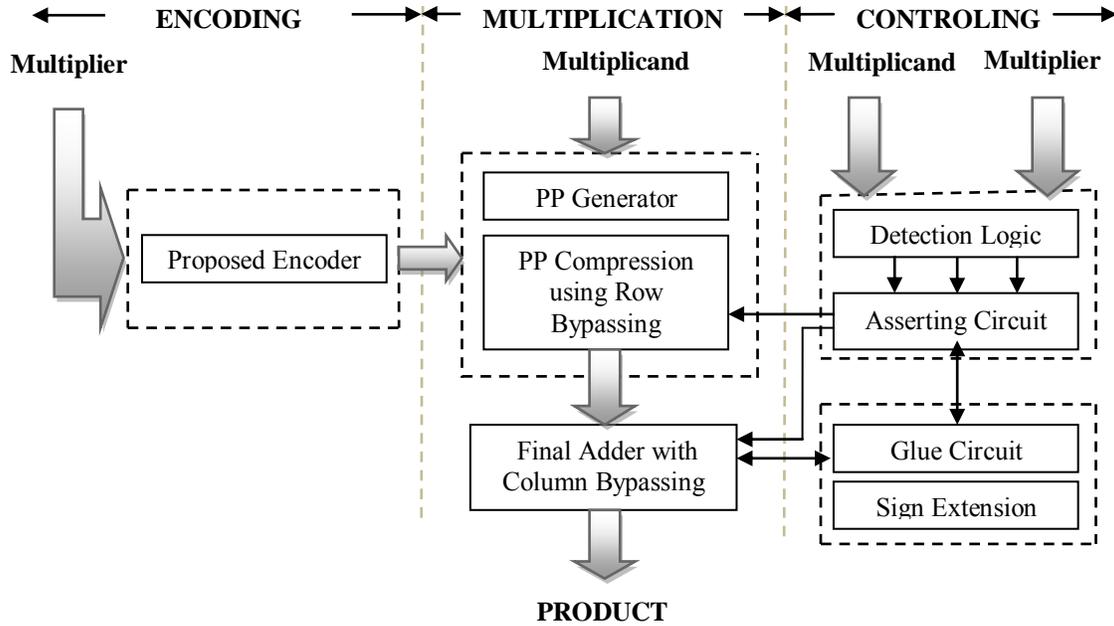

Figure 4. Block diagram of proposed hybrid encoded low power multiplier

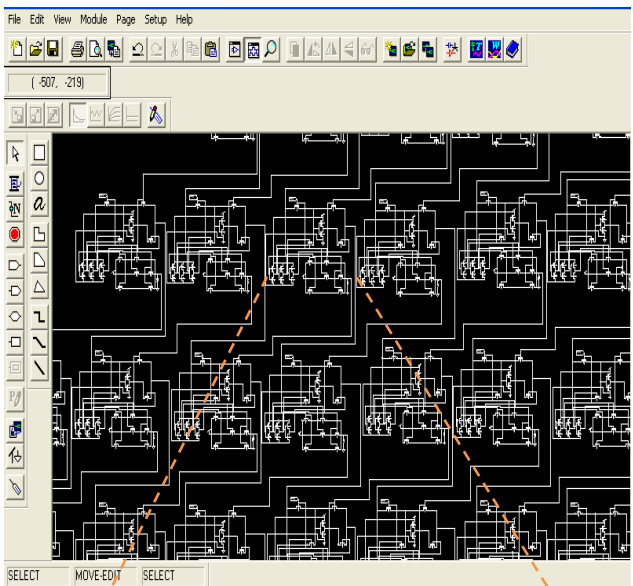

Figure 5. Architecture of multiplier with proposed adders

The output of the EDA editor is shown in Fig 5. The simulated adder is shown in enlarged version for understanding. The power analysis of the proposed multiplier-adder circuit has been estimated with example. For multiplying 65, which is the pixel value in Multiply and Accumulate unit (MAC) with another pixel value 34, the proposed procedure shown in Fig 6 may be adopted.

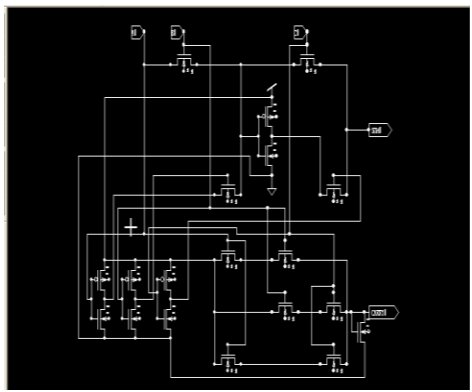

| Multiplicand (65) | 0 1 0 0 0 0 0 1 |
| Multiplier (34) | X 0 0 1 0 0 0 1 0 0 |

Booth recoding (4PP)   +1  -2  +1  -2

Proposed recoding (1PP)   Category D

Result (2210)   1 0 0 0 1 0 1 0 0 0 1 0

Figure 6. Hybrid encoded multiplication scheme for the pixel value multiplication

For the above multiplication, it needs 8 partial products for normal multiplication and 4 partial products for Booth recoding but only one partial product is enough for the proposed hybrid encoding method. Moreover, the proposed technique doesn't need the 2's complement process and virtual 0 which is to be placed as a first bit of Booth recoding. Table II shows the power and delay analysis of different multipliers for the stated multiplication. The simulation results have been taken for different voltage ranges from 0.8V to 2.4V. The power consumption of the proposed multiplier has been reduced by 87% and 26% compared with conventional and Booth multiplier.

TABLE II.  POWER AND DELAY ANALYSIS OF DIFFERENT MULTIPLIERS

| Multiplier type | Parameter | VDD (volts) | | | | | | | | |
|---|---|---|---|---|---|---|---|---|---|---|
| | | *0.8* | *1.0* | *1.2* | *1.4* | *1.6* | *1.8* | *2.0* | *2.2* | *2.4* |
| Conventional multiplier (8 PP) [6] | Power (µw) | 31.98 | 84.56 | 122.5 | 167.8 | 246.9 | 413.7 | 525.0 | 625.59 | 662.2 |
| | Delay (ns) | 11.20 | 5.138 | 4.165 | 3.213 | 2.765 | 2.443 | 2.296 | 2.1910 | 1.932 |
| Booth multiplier (4 PP) [5] | Power (µw) | 13.71 | 36.24 | 52.50 | 69.75 | 105.8 | 177.3 | 225.0 | 268.11 | 283.8 |
| | Delay (ns) | 4.800 | 2.200 | 1.790 | 1.380 | 1.190 | 1.050 | 0.980 | 0.9400 | 0.830 |
| Proposed multiplier (1 PP) | Power (µw) | 4.569 | 12.08 | 17.50 | 23.25 | 35.27 | 59.10 | 75.00 | 89.370 | 94.60 |
| | Delay (ns) | 1.600 | 0.734 | 0.595 | 0.459 | 0.395 | 0.349 | 0.328 | 0.3130 | 0.276 |

The full adder cell which is the important sub module of the proposed multiplier architecture is designed according to the following equations.

$$C = AB + BC + CA \qquad (1)$$
$$S = A'B'C + A'BC' + AB'C' + ABC \qquad (2)$$

In full adder, four inverters can be used to provide inverted inputs, the sum and carry circuits are joined together. A pull down nMOS transistor is connected near the carry output to provide the undistorted output. The output wave form of the full adder without and with the pull down transistor is shown in Fig 7 and Fig 8 respectively. Here 0.13µm TSMC technology files were used, for simulating in TANNER 12.6 EDA tool.

The various adder circuits have been simulated using the TSPICE TANNER 12.6 EDA tool for supply voltages range from 0.8V to 2.4 V. The operating frequency is set at 100 MHz. The power consumption variation with various voltages is shown in Fig 9.

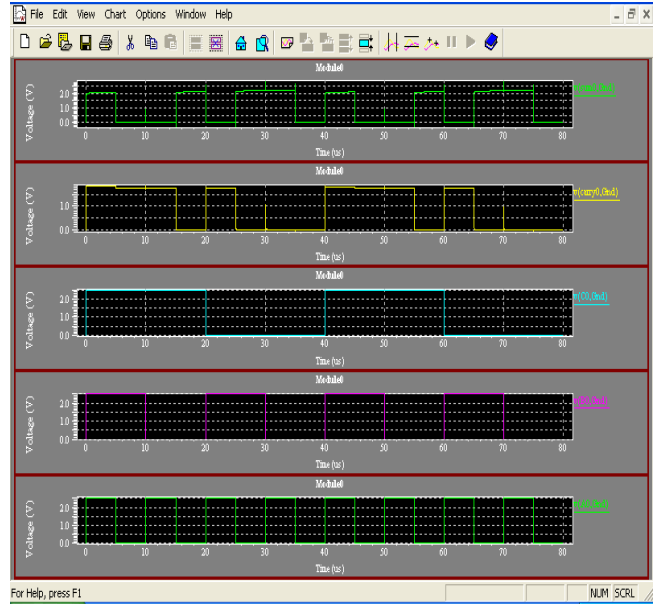

Figure 8. Output waveform of the full adder with pull down mechanism

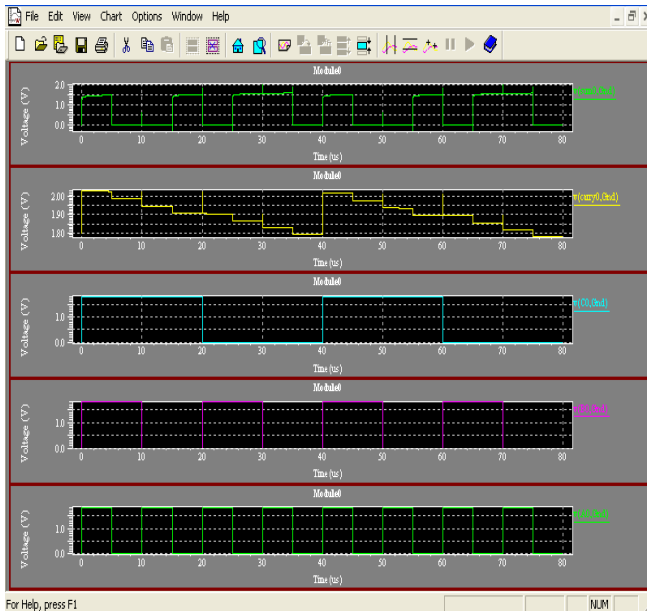

Figure 7. Output waveform of the full adder without pull down mechanism.

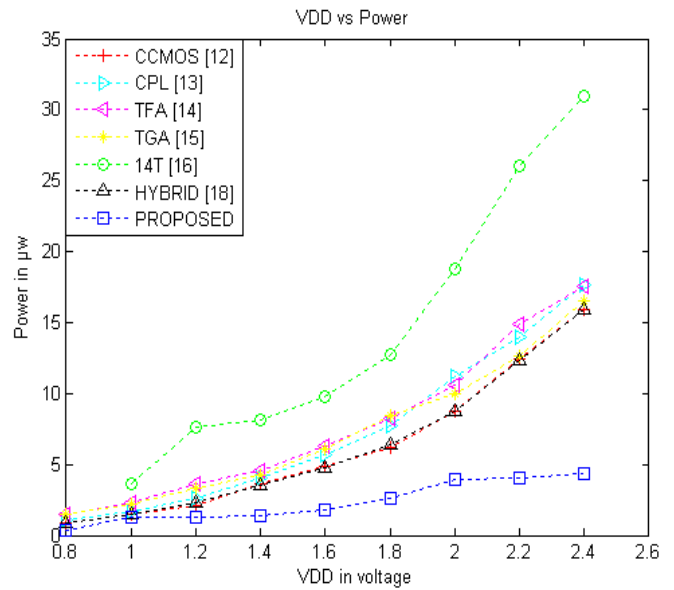

Figure 9. Variation of power consumption with input for different adders.

It is seen from the figure that the 14T design consumes more power beyond the supply voltage range 0.8V. All other designs C-CMOS, TGA, TFA, CPL, hybrid and the proposed method are working better at the input supply voltage ranges from 0.8V to 2.4V. Even the number of transistors required to design TGA and TFA is less, they require additional buffers at the output. This additional buffer increases the short circuit power and also switching power because of less driving capacity. CPL adder design consumes more power than hybrid and C-CMOS due to its dual-rail structure and the large number of internal nodes. Even though the transistor count of the proposed adder design is more than the 10T and 14T, the proposed adder cell consumes less power than other design which is shown in the comparison.

## IV. CONCLUSION

The performance of the SSST equipped low power multiplier with hybrid encoding has been estimated and compared with existing multipliers. The proposed encoding technique reduces the number of spurious switching activity and dynamic power consumption by analyzing the bit patterns in the input data. A low power full adder cell, which consumes less power than the other adders has been used to design the proposed multiplier. The switching activity of the proposed multiplier has been reduced by 86% and 46% compared with conventional and Booth multiplier respectively. It is observed from the device level simulation using TANNER 12.6 EDA that the power consumption of the proposed multiplier can be reduced by 87% and 26% compared with conventional and Booth multiplier.

AUTHORS PROFILE

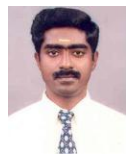

S.Saravanan received his B.E. Degree in Electrical and Electronics Engineering from Madras University, Tamilnadu, India in 2000 and M.E. Degree in Applied Electronics from Anna University, Tamilnadu, India in 2005. He is currently working towards the Ph.D degree in Information and Communication Engineering in Anna University, Chennai. Tamilnadu, India and working as an Assistant Professor in ECE Department, K.S.Rangasamy college of Technology, Tamilnadu, India.

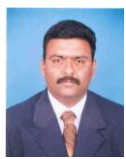

Dr. M. Madheswaran has obtained his Ph.D. degree in Electronics Engineering from Institute of Technology, Banaras Hindu University, Varanasi in 1999 and M.E degree in Microwave Engineering from Birla Institute of Technology, Ranchi, India. He has started his teaching profession in the year 1991 to serve his parent Institution Mohd. Sathak Engineering College, Kilakarai where he obtained his Bachelor Degree in ECE. He has served KSR college of Technology from 1999 to 2001 and PSNA College of Engineering and Technology, Dindigul from 2001 to 2006. He has been awarded Young Scientist Fellowship by the Tamil Nadu State Council for Science and Technology and Senior Research Fellowship by Council for Scientific and Industrial Research, New Delhi in the year 1994 and 1996 respectively. His research project entitled "Analysis and simulation of OEIC receivers for tera optical networks" has been funded by the SERC Division, Department of Science and Technology, Ministry of Information Technology under the Fast track proposal for Young Scientist in 2004. He has published 120 research papers in International and National Journals as well as conferences. He has been the IEEE student branch counselor at Mohamed Sathak Engineering College, Kilakarai during 1993-1998 and PSNA College of Engineering and Technology, Dindigul during 2003-2006. He has been awarded Best Citizen of India award in the year 2005 and his name is included in the Marquis Who's Who in Science and Engineering, 2006-2007 which distinguishes him as one of the leading professionals in the world. His field of interest includes semiconductor devices, microwave electronics, optoelectronics and signal processing.  He is a member of IEEE, SPIE, IETE, ISTE, VLSI Society of India and Institution of Engineers (India).